# Diffusion Model for Interest Refinement in Multi-Interest Recommendation

Yankun Le*
Northeastern University, China
2401851@stu.neu.edu.com

Haoran Li*
Xiaohongshu
haoran.li.cs@gmail.com

Baoyuan Ou
Xiaohongshu
oubaoyuan@xiaohongshu.com

Yingjie Qin
Xiaohongshu
yingjieqin@xiaohongshu.com

Zhixuan Yang
Northeastern University, China
2472144@stu.neu.edu.cn

Ruilong Su
Xiaohongshu
suruilong@xiaohongshu.com

Fu Zhang†
Northeastern University, China
zhangfu216@126.com

## Abstract

Multi-interest candidate matching plays a pivotal role in personalized recommender systems, as it captures diverse user interests from their historical behaviors. Most existing methods utilize attention mechanisms to generate interest representations by aggregating historical item embeddings. However, these methods only capture overall item-level relevance, leading to coarse-grained interest representations that include irrelevant information. To address this issue, we propose the Diffusion Multi-Interest model (DMI), a novel framework for refining user interest representations at the dimension level. Specifically, DMI first introduces controllable noise into coarse-grained interest representations at the dimensional level. Then, in the iterative reconstruction process, DMI combines a cross-attention mechanism and an item pruning strategy to reconstruct the personalized interest vectors with the guidance of tailored collaborative information. Extensive experiments demonstrate the effectiveness of DMI, surpassing state-of-the-art methods on offline evaluations and an online A/B test. Successfully deployed in the real-world recommender system, DMI effectively enhances user satisfaction and system performance at scale, serving the major traffic of hundreds of millions of daily active users. [1]

*Both authors contributed equally to this research.
†Corresponding author.

[1]The code will be released for reproducibility once the paper is accepted.

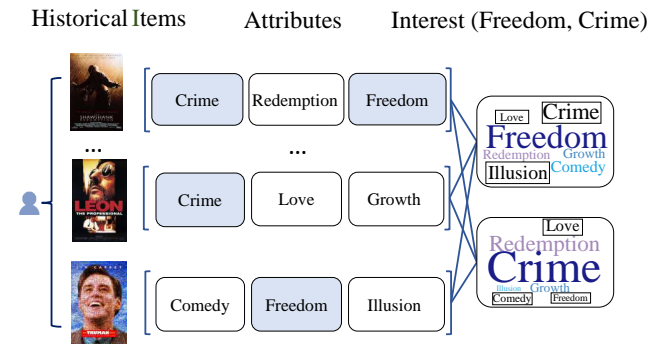

Figure 1: Each user's historical behaviors include multiple attributes, but only a subset of these attributes align well with the user's specific interests. Current methods extract interest features at the item level, failing to focus on interest-relevant attributes effectively. As a result, irrelevant attributes is included in the final interest representations, as highlighted in the black box.

## 1 INTRODUCTION

To deliver personalized recommendation lists, industrial recommendation systems depend on candidate matching [16] to filter a vast number of products on online platforms. This process focuses on retrieving a small subset of items from the item corpus that accurately captures user interests while maintaining low computational costs. As a critical component of recommender systems, candidate matching plays a decisive role in determining the overall system's performance.

By extracting multiple interest vectors from user behavior sequences, the multi-interest representation framework significantly improves the diversity of candidate sets in the matching stage, gaining notable recognition within the recommendation systems community. MIND [11] was the first to capture users' multiple interests using dynamic routing within Capsule Networks [21]. Building on this, ComiRec [1] introduces multi-head attention to better encode diverse user interests, thereby enhancing recommendation diversity. Subsequent works [2, 3] have incorporated additional elements, such as periodicity and user profiles, to further enhance interest modeling. Beyond these improvements, emerging approaches have introduced auxiliary loss functions to enhance the interest activation process. For example, Re4 [27] and TiMiRec [24] use auxiliary loss to distinguish and distill target interests. The recent work REMI [25] finds that current multi-interest models miss valuable insights by overemphasizing closely related historical data. To mitigate this, REMI introduces a regularization method that adjusts the contribution weights of historical behaviors. These methods typically use attention mechanisms to compute the relevance between user interests and item embeddings, which are used to weight item embeddings and derive interest-aware features. However, these methods only capture overall item-level relevance, leading to coarse-grained interest representations that include interest-irrelevant attributes. As shown in Figure 1, when a user expresses a personalized interest in a movie topic such as 'freedom,' the final aggregated representation may include irrelevant attributes like 'crime,' 'hallucination,' or 'love.' While these attributes may align with the user's marginal interests, they hinder the model's ability to accurately capture the user's primary interest in 'freedom.'

In recent years, diffusion models (DM) have drawn increasing attention in recommender systems due to their powerful denoising capabilities [9, 14, 28, 29] and ability to model complex distributions [17, 26]. These models progressively corrupt user intent representations by adding Gaussian noise in a controlled manner and then iteratively reconstruct the clean representation from the noisy version. This approach aligns well with the process of multi-interest learning in recommender systems, where aggregating interest-irrelevant attributes can be considered as adding noise. Subsequently, the reverse process progressively refines the representation by removing unrelated interest attributes from the noise distribution. In this process, DM acquires the ability to model the complex distribution of user-personalized interests, alleviating the limitations of item-level extraction. In light of these, it is promising to incorporate the diffusion model to enhance multi-interest representation learning of recommender models.

Therefore, we prsent the **D**iffusion **M**ulti-**I**nterest model (DMI), which aims to refine user interest vectors at the dimension-level. First, DMI extracts coarse-grained user interest vectors with a base multi-interest recommender. This module generates interest representations by aggregating item embeddings through attention mechanisms. Second, to obtain fine-grained interest vectors, we introduce a novel denosing module which comprises a forward and a reverse process. Specifically, in the forward process, we progressively introduce controllable Gaussian noise to simulate noisy user interests. Then, in the reverse process, DMI employs a cross-attention mechanism to reconstruct cleaner interest vectors with the guidance of tailored collaborative information. This information is constructed by our item pruning strategy, which prunes item embeddings with lower interest relevance, as determined by their corresponding attention weights. By iterative denoising, DMI obtains a highly personalized interest representation. To balance the learning of interest extraction and refinement, the diffusion reconstruction loss will only update the diffusion module while the recommendation loss will impact both the diffusion module and the multi-interest extractor.

We conduct extensive experiments on three public benchmarks. Various analyses including ablation study, hyper-parameter analysis, and result analysis validate the practical merits of DMI in capturing fine-grained interest representations. In summary, the main contributions of our work are threefold:

- We identify the limitations of existing multi-interest models: they only focus on item-level interest extraction, which would include irrelevant interest attributes.
- We propose a novel Diffusion Multi-Interest model (DMI) to address the limitations by refining interest vectors at the dimension level. DMI incorporates an item pruning strategy to construct tailored collaborative information and a cross-attention denoising module for interest-aware reconstruction, enabling more precise and personalized user interest modeling.
- Extensive offline evaluations and an online A/B test validate the effectiveness of our proposed DMI. Comparative analysis further demonstrates that DMI captures more refined user interests, retrieving more personalized and diverse items.

## 2 PRELIMINARIES

In this section, we will provide a brief overview of multi-interest candidate matching and diffusion models (DMs).

**Multi-Interest Candidate Matching.** Suppose we have a group of users denoted as $\mathcal{U}$ and a large item corpus as $\mathcal{I}$. Each user $u \in \mathcal{U}$ has a history of behaviors $(i_1, i_2, \ldots, i_n)$ sorted by time, where $i_n$ is the last item the user interacted with. In recommender systems, the primary objective of multi-interest candidate matching is to retrieve a subset of items from $\mathcal{I}$ that the user is likely to interact with in the future. To select the top-$N$ candidate items for user $u$, the similarity score can be calculated as:

$$s_{u,t} = \max_{1 \leq k \leq K} ((e_k^u)^T r_t) \quad (1)$$

where $s_{u,t}$ denotes the similarity score between user $u$ and target item $t$, $e_u^k$ denotes the $k$-th interest vector of user $u$, and $r_t$ is the representation of the target item $t$.

**Diffusion Models.** DMs have achieved remarkable results in Computer Vision and Natural Language Processing. They mainly consist of a forward and a reverse process. In the forward process, given an input sample $x_0 \sim q(x_0)$, the latent variables $\{x_1, ..., x_T\}$ are generated by a fixed Markov process with Gaussian transitions, which can be formulated as:

$$q(x_t|x_{t-1}) = \mathcal{N}(x_t; \sqrt{1-\beta_t}x_{t-1}, \beta_t \mathcal{I}) \quad (2)$$

where $\beta_t$ denotes the variance of the additive noise, $\mathcal{N}$ denotes the Gaussian distribution, and $t$ denotes the diffusion step.

In the reverse process, DMs remove added noise to recover $x_{t-1}$ from $x_t$, which can be formulated as:

$$p_\theta(x_{t-1}|x_t) = \mathcal{N}(x_{t-1}; \mu_\theta(x_t, t), \Sigma_\theta(x_t, t)) \quad (3)$$

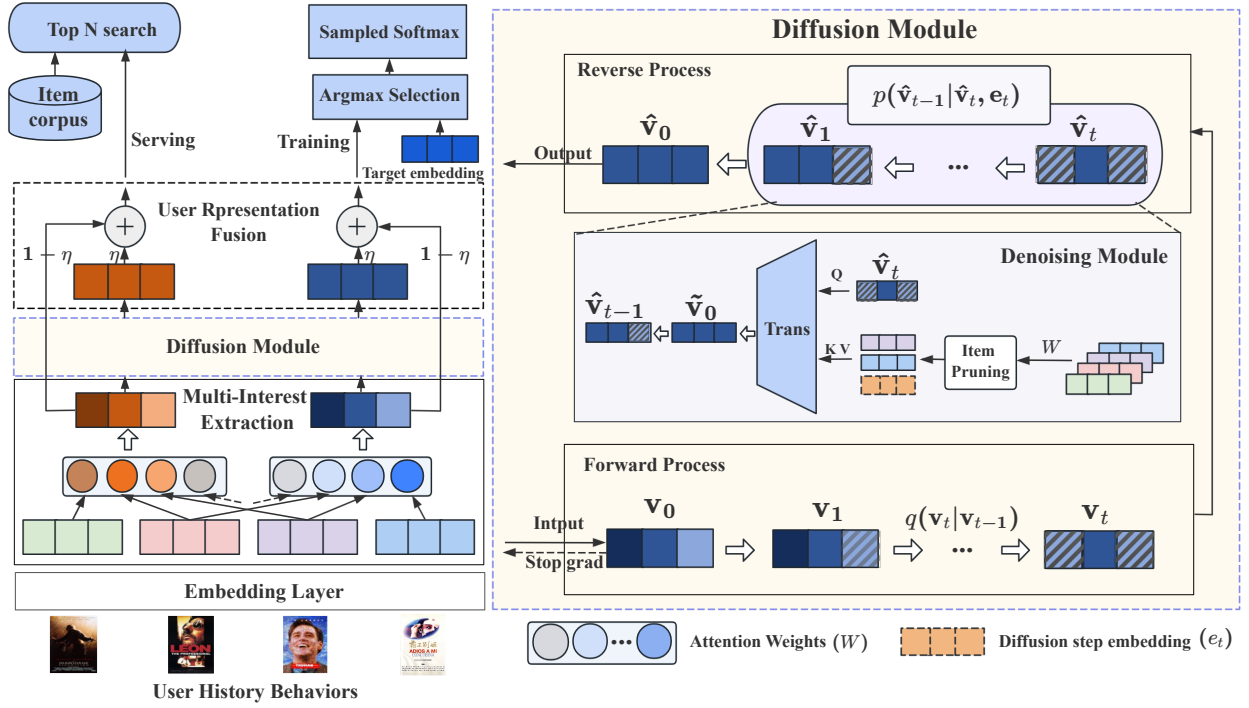

Figure 2: Overview of DMI. Given user history behaviors, the multi-interest extraction generates attention weights and derives interest vectors. Based on this, the diffusion module introduce controllable Gaussian noise to simulate noisy user interests in the forward process. Subsequently, in the reverse process, the diffusion module incorporates the item pruning strategy and a transformer layer to selectively reconstruct the interest-relevant parts.

where $\mu_\theta(x_t, t)$ and $\Sigma_\theta(x_t, t)$ are are the mean and covariance of the Gaussian distribution predicted by a neural network parameterized by $\theta$.

## 3 DMI: DIFFUSION MULTI-INTEREST MODEL

In this section, as illustrated in Figure 2, we will introduce DMI, a diffusion multi-interest model for enhancing the interest representations in recommender systems. Specifically, DMI consists of three complementary modules. First, we implement a multi-interest extraction module to capture diverse user interests from their historical items. Second, we design a diffusion module to inject noise into the interest vectors and reconstruct a cleaner version, aiming to refine user interest representations at the dimension level. Lastly, we develop an item pruning strategy to adaptively guide the denoising module, which incorporates the cross-attention mechanism for interest-aware reconstruction. Next, we will detail these three key modules in detail as follows:

### 3.1 Multi-Interest Extraction

In DMI, we implement a base multi-interest recommender to capture diverse user interests. Specifically, we employ a self-attentive method, as in ComiRec-SA[1], to aggregate interest information from historical items.

Formally, given the embeddings of user historical items, $H \in \mathbb{R}^{T \times d}$, where $T$ denotes historical sequence length and $d$ denotes the dimension of embeddings, we use the self-attention mechanism to obtain the attention weights $A \in \mathbb{R}^{K \times T}$:

$$A = \text{softmax}\left(W_2 \tanh\left(W_1 H^\top\right)\right) \qquad (4)$$

where $W_1 \in \mathbb{R}^{d_a \times d}$ and $W_2 \in \mathbb{R}^{K \times d_a}$ are trainable parameters, and $K$ denotes the number of interest vectors.

Then, we can obtain interest vectors $V \in \mathbb{R}^{K \times d}$ by summing up item embeddings according to the attention weights:

$$V = AH \qquad (5)$$

However, this approach indiscriminately considers all historical items, leading to potential noise in the final interest vectors. Therefore, we propose a novel denosing module to derive cleaner and more accurate interest representations. In the following sections, all operations are performed on one selected interest vector $v$. This principle holds consistency between the training and inference processes: in the training process, the interest vector most similar to the target embedding is selected (Sec. 3.3 for details), while in the inference stage, all interest vectors are sequentially used for retrieval.

## 3.2 Diffusion for Interest Refinement

After obtaining user interest vector **v** from the multi-interest extraction module, it is likely that it includes interest-irrelevant information in some dimensions, as shown in Figure 1. This significantly affects the accuracy of personalized recommendations. To address this issue, we propose a diffusion module to iteratively refine the interest embeddings in recommender systems.

Specifically, the diffusion module consists of two processes: a forward and a reverse process, as shown in the right part of Figure 2. In the forward process, scheduled Gaussian noise is injected into the interest vector **v** to add noise. In the reverse process, DMI iteratively reconstructs a cleaner representation with a denoising module. To guide the denoising module in differentiating between interest-relevant and irrelevant attributes, we propose an item pruning strategy together with a cross-attention mechanism that adaptively integrate user interest information in the reconstruction process.

**Forward Process.** We denote the selected interest vector $\mathbf{v} \in \mathbb{R}^{1 \times d}$ obtained from the multi-interest extraction module as the initial state $\mathbf{v}_0$. In the forward process, the latent variables $\{\mathbf{v}_1, ..., \mathbf{v}_T\}$ are generated by a fixed Markov process with Gaussian transitions, which can be formulated as:

$$q(\mathbf{v}_t \mid \mathbf{v}_{t-1}) = \mathcal{N}(\mathbf{v}_t; \sqrt{1 - \beta_t}\mathbf{v}_{t-1}, \beta_t \mathcal{I}) \quad (6)$$

where $\beta_t$ denotes the variance of the additive noise at each step $t$.

By employing the reparameterization trick [22], we can directly obtain $\mathbf{v}_t$ from $\mathbf{v}_0$:

$$q(\mathbf{v}_t \mid \mathbf{v}_0) = \mathcal{N}(\mathbf{v}_t; \sqrt{\bar{\alpha}_t}\mathbf{v}_0, (1 - \bar{\alpha}_t)I) \quad (7)$$

where $\alpha_t = 1 - \beta_t$, $\bar{\alpha}_t = \prod_{t'=1}^{t} \alpha_{t'}$. Let $\varepsilon \sim \mathcal{N}(0, I)$, at any step $t$, $\mathbf{v}_t$ can be sampled as $\mathbf{v}_t = \sqrt{\bar{\alpha}_t}\mathbf{v}_0 + \sqrt{1 - \bar{\alpha}_t}\varepsilon$.

Following the [28], we use the linear variance noise schedule to regulate the noise lever at each step:

$$1 - \bar{\alpha}_t = s \cdot \left[ \alpha_{\min} + \frac{t-1}{T-1}(\alpha_{\max} - \alpha_{\min}) \right] \quad (8)$$

where $\alpha_{\min}$ and $\alpha_{\max}$ are predefined parameters that determine the minimum and maximum noise levels at each step. To preserve the personalized information in the interest vectors, we reduce the added noises (compared to the CV domain [18]) and avoid corrupting users' interactions into pure noises. Additionally, this strategy helps avoid starting from pure noise during inference. [17, 28, 29].

**Reverse Process.** To reconstruct cleaner interest vector from $\mathbf{v}_T$, the denoising module iteratively recovers users' interactions. It is worth noting that this process differs from the image generation domain, where the noise added at step $t$ is the target to be predicted. In contrast, in the recommendation domain, we use the denoising module to directly generate the denoised embedding from the previous step [30]. In summary, the reverse process at each step $t$ can be formulated as:

$$p_\theta(\hat{\mathbf{v}}_{t-1} \mid \hat{\mathbf{v}}_t) = \mathcal{N}(\hat{\mathbf{v}}_t; \mu_\theta(\hat{\mathbf{v}}_t, t), \Sigma_\theta(\hat{\mathbf{v}}_t, t)) \quad (9)$$

Following the standard DM method [18], we ignore the learning of $\Sigma_\theta(\hat{\mathbf{V}}_t, t)$, and directly set $\Sigma_\theta(\hat{\mathbf{V}}_t, t) = \sigma^2(t)I$ for simplicity and training stability. As for $\mu_\theta(\hat{\mathbf{V}}_t, t)$, in typical practice [29], it is commonly transformed into the following form by applying Bayes' theorem:

$$\mu_\theta(\hat{\mathbf{v}}_t, t) = \frac{\sqrt{\alpha_t}(1 - \bar{\alpha}_{t-1})}{1 - \bar{\alpha}_t}\hat{\mathbf{v}}_t + \frac{\sqrt{\bar{\alpha}_{t-1}}(1 - \alpha_t)}{1 - \bar{\alpha}_t}\tilde{\mathbf{v}}_0 \quad (10)$$

where $\tilde{\mathbf{v}}_0$ is the predicted $\mathbf{v}_0$ since $\mathbf{v}_0$ is unknown in the reverse precess compared to the forward process. To address this, the denoising module $f_\theta$ is specifically designed to estimate $\mathbf{v}_0$, which will be further illustrated in the next section.

**Denoising Module with Item Pruning.** To address the interest-irrelevant attributes issue, we propose a novel denoising module, which combines an item pruning strategy and a cross-attention mechanism to reconstruct user interest embeddings $\tilde{\mathbf{v}}_0$. Specifically, for noisy interest embedding $\mathbf{v}_t$ obtained from the forward process, the denoising module includes: (1) a cross-attention mechanism (implemented by a transformer layer) to incorporate user-specific interest information from historical behaviors; (2) an item pruning strategy that adaptively filters out interest-irrelevant information.

The denoising module $f_\theta$ can be formulated as follows:

$$\tilde{v}_0 = \textbf{Transformer}\,(Q = \hat{\mathbf{v}}_t, K = V = \text{concat}(\mathbf{e}_t, \mathbf{C}, \text{axis} = 1)) \quad (11)$$

where $\mathbf{e}_t$ denotes the diffusion step embedding and $\mathbf{C}$ denotes the top $k$ relevant historical item embeddings selected by our item-pruning strategy.

Concretely, introducing all historical behavior information **H** for reconstruction is not conducive to removing interest-irrelevant attributes. Therefore, we propose a simple and effective item pruning strategy to select the TopK most relevant historical item embeddings from **H**:

$$\mathbf{C} = \mathbf{H}[:, \underset{K=\gamma*n}{\text{TopK}}(\boldsymbol{a})] \quad (12)$$

where $\gamma \in (0, 1)$ controls the selected proportion. Here, $n$ represents the actual number of historical behavior sequences for each user, the TopK operation returns the indices of the top-$k$ values, and $\boldsymbol{a}$ denotes the attention weights corresponding to the selected interest vector.

During diffusion model training, the goal is to enhance reconstruction performance, ensuring the predicted representation closely matches the initial one. However, allowing the gradients of reconstruction loss to affect the multi-interest extraction module risks losing its intrinsic information by making it easier to reconstruct. To prevent this, we freeze the original interest vector **v** in the denoising module. In addition, to optimize the extractor by recommendation loss, we define the final user representation $\mathbf{Z}^u \in \mathbb{R}^{1 \times d}$ as:

$$\mathbf{Z}^u = \eta \cdot \tilde{\mathbf{v}}_0 + (1 - \eta) \cdot \mathbf{v} \quad (13)$$

where $1 - \eta$ denotes the weight of the initial coarse-grained interest vector. This operation allows gradients of the recommendation loss to update the multi-interest recommender.

## 3.3 DMI Training and Inference

**Training Stage.** During training, we utilize the final user representation $\mathbf{Z}^u$ to compute the sampled softmax loss $\mathcal{L}_s$ for the recommendation task, while $\tilde{v}$ is employed to calculate the Euclidean distance $\mathcal{L}_{dm}$ as the reconstruction loss for the denoising module.

Specifically, for the multi-interest extraction module, given the user interest vectors $\mathbf{V}$ and the target item representation $r_t$, we apply an argmax operation to activate a specific user interest $\mathbf{v} \in \mathbb{R}^d$ from $\mathbf{V} \in \mathbb{R}^{K \times d}$:

$$\mathbf{v} = \mathbf{V}\left[\operatorname{argmax}(\mathbf{V}^\top r_t)\right] \quad (14)$$

After interest refinement, we obtain the final user representation $\mathbf{Z}^u$. To maximize the probability of the user $u$ interacting with the target item $r_t$ in the training stage, we utilize the sampled softmax method [25] to calculate the loss:

$$P_\theta(t|u) = \frac{\exp(\mathbf{Z}^{u\top} r_t)}{\sum_{r_k \in \mathcal{I}} \exp(\mathbf{Z}^{u\top} r_k)} \quad (15)$$

$$\mathcal{L}_S = \sum_{u \in \mathcal{U}} \sum_{i_t \in \mathcal{I}_u} -\log P_\theta(i_t|u) \quad (16)$$

For the diffusion module, the primary objective is to drive the posterior distribution $p_\theta(\mathbf{v}_{t-1} \mid \mathbf{v}_t)$ to approximate the prior distribution $q(\mathbf{v}_t \mid \mathbf{v}_{t-1})$ during the reverse process. In other words, this process minimizes the variational bound between the two distributions, with the objective function given as:

$$\mathcal{L}_{vlb} = D_{KL}\left(q(\mathbf{v}_{t-1} \mid \mathbf{v}_t, \mathbf{v}_0) \parallel p_\theta(\mathbf{v}_{t-1} \mid \mathbf{v}_t)\right) \quad (17)$$

Through Bayesian inference, we can deduce that the key to bringing these two distributions closer lies in reducing the gap between their means. Essentially, this implies that optimizing our denoising module requires minimizing the distance between $v_0$ and $f_\theta(v_t, t, C)$. In practice, we uniformly sample $t$ from $\{1, 2, \ldots, T\}$ to simulate denoising across various noise levels. Considering that mean squared error (MSE) can be directly used as the optimization objective [18], in our work, we adopt Euclidean distance as the optimization target to further penalize distributional differences.

$$\mathcal{L}_{dm} = \sqrt{\sum_{i=1}^{d} (\tilde{\mathbf{v}}_0 - \mathbf{v}_0)^2} \quad (18)$$

Here, $\mathbf{v}_0$ represents the interest vector selected via argmax. The training procedure of DMI is stated in Algorithm 1.

Overall, the entire training objective can be formulated as:

$$\mathcal{L} = \mathcal{L}_S + \lambda \mathcal{L}_{dm} \quad (19)$$

where $\lambda$ is the hyperparameter to balance the two training objectives for better convergence.

**Inference Stage.** During the inference stage, we sequentially select the interest vectors for interest refinement. In the forward process, the sampled diffusion step is set to $T$ to align with the training stage. Then in the reverse process, DMI iteratively refines $\mathbf{v}_T$ for $T$ steps. After obtaining the refined interest vectors $\hat{\mathbf{v}}_0$, we use $\hat{\mathbf{v}}_0$ instead of $\tilde{\mathbf{v}}_0$ to compute the final user representation $\mathbf{Z}^u$ and retrieve the top-$N$ items from $\mathcal{I}$. To aggregate the final user recommendation list, we rank and filter the items retrieved based on their similarity with each final user representation $\mathbf{Z}^u_k$, which can be calculated as:

$$f(u, i) = \max_{1 \le k \le K} \left((\mathbf{Z}^u_k)^\top \mathbf{H}_i\right) \quad (20)$$

where $\mathbf{H}_i$ represents the embedding of the item $i \in \mathcal{I}$. The inference procedure of DMI is stated in Algorithm 2.

---

**Algorithm 1** DMI Training

**Input:** all users $\bar{U}$, diffusion step $T$, denoising module $f_\theta$
1: Sample a batch of users $U \subset \bar{U}$.
2: **for all** $u \in U$ **do**
3:     Select the most similar interest vector $\mathbf{v}$ via Eq. (14);
4:     Sample $t \sim \mathcal{U}(1, T)$
5:     Compute $\mathbf{v}_t$ given $\mathbf{v}_0$ and $T$ via $q(\mathbf{v}_T|\mathbf{v}_0)$ in Eq. (7);
6:
7:     Reconstruct $\tilde{\mathbf{v}}_0$ through the denoising module $f_\theta$;
8:     Calculate $\mathcal{L}_{dm}$ by Eq. (18);
9:     Compute final user representations $\mathbf{Z}^u$ given $\tilde{\mathbf{v}}_0$ and $\mathbf{v}$;
10:    Calculate $\mathcal{L}_S$ by Eq. (16);
11:    Calculate $\mathcal{L}$ by Eq. (19);
12:
13:    Take gradient descent step on $\nabla_\theta(\mathcal{L})$ to optimize $\theta$;
14: **end for**
**Output:** optimized $\theta$

---

**Algorithm 2** DMI Inference

**Input:** all users $\bar{U}$, diffusion step $T$, denoising module $f_\theta$
1: Sample a batch of users $U \subset \bar{U}$.
2: **for all** $u \in U$ **do**
3:     **for all** $\mathbf{v} \in V$ **do**
4:         Compute $\mathbf{v}_T$ given $\mathbf{v}_0$ and $T$ via $q(\mathbf{v}_T|\mathbf{v}_0)$ in Eq. (7);
5:         **for** $t = T, \ldots, 1$ **do**
6:             Reconstruct $\tilde{\mathbf{v}}_0$ through $f_\theta$;
7:             Compute $\hat{\mathbf{v}}_{t-1}$ from $\hat{\mathbf{v}}_t$ and $\tilde{\mathbf{v}}_0$ via Eq. (9) and (10);
8:         **end for**
9:         Compute final user representations $\mathbf{Z}^u$ given $\hat{\mathbf{v}}_0$ and $\mathbf{v}$;
10:        Retrieve top-N items from the item pool by Faiss
11:     **end for**
12:     Determine the overall top-N item candidates via Eq. (20);
13: **end for**
**Output:** the overall top-N item candidates

---

| Dataset | #Users | #Items | #Interactions | Density |
|---|---|---|---|---|
| Book | 603,668 | 367,982 | 8,898,041 | 0.004006% |
| Beauty | 22,363 | 12,101 | 198,502 | 0.07335% |
| Gowalla | 29,858 | 40,981 | 1,027,370 | 0.0840% |

**Table 1: Statistics of datasets.**

## 4 EXPERIMENTS

In this section, we conduct experiments on three large-scale real-world datasets to answer the following research questions:
- **RQ1:** How does DMI perform compared to state-of-the-art models?
- **RQ2:** Can the proposed modules (e.g., denoising module) effectively improve performance?
- **RQ3:** How effective is DMI in generating more fine-grained and focused interest representations?
- **RQ4:** How does DMI perform under different hyperparameter settings?

Table 2: Performance of different methods on three large-scale datasets. We report the Recall (R), Hit Rate (HR), and Normalized Discounted Cumulative Gain (ND). The best results and second best results are bold and underlined, respectively.

| Dataset | Metric | PoP | GRU4Rec | Y-DNN | MIND | ComiRec | Re4 | PIMIRec | REMI | DMI | Improv. |
|---|---|---|---|---|---|---|---|---|---|---|---|
| Book | R@20 | 0.0158 | 0.0441 | 0.0467 | 0.0420 | 0.0557 | 0.0597 | 0.0682 | <u>0.0826</u> | **0.0942**± **0.0013** | +14.4% |
|  | HR@20 | 0.0345 | 0.1004 | 0.1043 | 0.0986 | 0.1142 | 0.1240 | 0.1411 | <u>0.1650</u> | **0.1867**± **0.0021** | +13.1% |
|  | ND@20 | 0.0143 | 0.0378 | 0.0391 | 0.0357 | 0.0446 | 0.0476 | 0.0526 | <u>0.0623</u> | **0.0699**± **0.0004** | +12.1% |
|  | R@50 | 0.0281 | 0.0706 | 0.0722 | 0.0687 | 0.0863 | 0.0690 | 0.1056 | <u>0.1189</u> | **0.1402**± **0.0008** | +17.9% |
|  | HR@50 | 0.0602 | 0.1553 | 0.1607 | 0.1533 | 0.1796 | 0.1975 | 0.2062 | <u>0.2298</u> | **0.2656**± **0.0022** | +15.5% |
|  | ND@50 | 0.0193 | 0.0443 | 0.0457 | 0.0433 | 0.0511 | 0.0576 | 0.0583 | <u>0.0657</u> | **0.0751**± **0.0003** | +14.3% |
| Beauty | R@20 | 0.0212 | 0.0421 | 0.0596 | 0.0678 | 0.0581 | 0.0652 | 0.0658 | <u>0.0876</u> | **0.1018**± **0.0022** | +16.2% |
|  | HR@20 | 0.0407 | 0.0784 | 0.1065 | 0.1238 | 0.1046 | 0.1186 | 0.1205 | <u>0.1537</u> | **0.1821**± **0.0008** | +18.4% |
|  | ND@20 | 0.0152 | 0.0376 | 0.0423 | 0.0465 | 0.0396 | 0.0441 | 0.0417 | <u>0.0591</u> | **0.0662**± **0.0012** | +12.01% |
|  | R@50 | 0.0426 | 0.0683 | 0.0895 | 0.1146 | 0.1043 | 0.1063 | 0.1027 | <u>0.1448</u> | **0.1631**± **0.0018** | +12.6% |
|  | HR@50 | 0.0813 | 0.1297 | 0.1680 | 0.1913 | 0.1699 | 0.1785 | 0.2073 | <u>0.2369</u> | **0.2641**± **0.0025** | +11.5% |
|  | ND@50 | 0.0248 | 0.0438 | 0.0537 | 0.0557 | 0.0501 | 0.0528 | 0.0517 | <u>0.0680</u> | **0.0751**± **0.0009** | +10.4% |
| Gowalla | R@20 | 0.0037 | 0.1119 | 0.1191 | 0.1203 | 0.1021 | 0.0843 | 0.1396 | <u>0.1460</u> | **0.1648**± **0.0031** | +12.8% |
|  | HR@20 | 0.0112 | 0.4164 | 0.4449 | 0.4475 | 0.4027 | 0.3104 | 0.4874 | <u>0.4998</u> | **0.5415**± **0.0073** | +8.3% |
|  | ND@20 | 0.0081 | 0.1687 | 0.1746 | 0.1779 | 0.1518 | 0.1287 | 0.1910 | <u>0.2003</u> | **0.2156**± **0.0017** | +7.5% |
|  | R@50 | 0.0063 | 0.1929 | 0.2057 | 0.2102 | 0.1910 | 0.1403 | 0.2426 | <u>0.2487</u> | **0.2735**± **0.0020** | +9.9% |
|  | HR@50 | 0.0241 | 0.5848 | 0.6216 | 0.6109 | 0.5088 | 0.4224 | 0.6749 | <u>0.6809</u> | **0.7180**± **0.0137** | +5.4% |
|  | ND@50 | 0.0147 | 0.1820 | 0.1913 | 0.1897 | 0.1731 | 0.1410 | 0.2086 | <u>0.2116</u> | **0.2263**± **0.0032** | +6.9% |

- **RQ5:** How effective is DMI in real-world recommender systems?

## 4.1 Experimental Settings

*4.1.1 Datasets.* We evaluate the effectiveness of DMI on three large-scale public datasets, i.e., Amazon Book, Amazon Beauty, and Gowalla.
- **Amazon Book and Beauty** [2] consist of product view data from the widely-used Amazon platform. For evaluation, we select the largest subset from the Book category and a smaller, yet denser subset from the Beauty category. The maximum sequence length is set to 20.
- **Gowalla** [4] is a typical check-in dataset built from a location-based social networking website. The maximum sequence length is set to 40.

The dataset preprocessing follows the prior study [25]. We set the filter size to 5 in the Amazon datasets which means removing all items and users that occur less than five times, and set the filter size to 10 in the Gowalla dataset. The processed datasets' statistics are provided in Table 1.

*4.1.2 Training and Evaluation Setup.* Following [25], we set the split ratio of training, validation, and test to 8:1:1. Our proposed solution is evaluated using several widely recognized metrics, including Recall, Hit Rate, and NDCG (Normalized Discounted Cumulative Gain). The metrics are computed with the top 20 and 50 matched candidates.

*4.1.3 Baseline Models.* The proposed DMI is compared with the following representative methods for candidate matching:
- **PopRec** is a traditional method that recommends the most prevalent items to users.

[2]http://jmcauley.ucsd.edu/data/amazon/

- **Y-DNN** [5] constructs a DNN model to obtain user representations, where embeddings are averaged and concatenated before feeding into MLP layers.
- **GRU4Rec** [7] is a session-based recommender that employs gated recurrent units (GRUs) to model temporal information between historical items.
- **MIND** [11] is designed to capture users' multiple interests and dynamically route their preferences for effective recommendation.
- **ComiRec-SA** [1] uses self-attention mechanisms to extract multiple user interests from behavior sequences, while an aggregation module balances the accuracy and diversity of its recommendations.
- **PIMIRec** [3] integrates periodicity and interactivity into a multi-interest framework for sequential recommendations, improving the model's ability to capture users' evolving interests and interactions over time.
- **Re4** [27] addresses the challenge of learning relevant user interests by re-contrasting, re-attending, and re-constructing user-item interactions.
- **REMI** [25] improves the training efficiency and mitigates routing collapse by implementing a Routing Regularization (RR) strategy, effectively distributing attention weight among relevant items.

*4.1.4 Implementation Details.* Since REMI results in a more uniform distribution of attention weight, we combine DMI with REMI as our default model setting. During experiments, we set the embedding dimension to 64, the batch size to 128, and the maximum number to 1 million. To remain fair comparison, the negative training sample size within every batch is set to 128 × 10. Following mainstream work settings, we set the interest number $k$ to 4. We use the Adam optimizer [10] with a learning rate of 0.002. The diffusion vector weight $\eta$ is set to 0.4 and the diffusion loss weight $\lambda$ is tuned within {3, 4, 5, 6, 7}. As for the hyperparameters that control the diffusion model, we fix $\alpha_{\min}$ to 0.0001 and adjust $\alpha_{\max}$ within the set {0.001, 0.002, 0.004, 0.008}. The diffusion step $T$ is empirically selected from the candidate set {5, 10, 20, 40, 80}.

## 4.2 Overall Performance (RQ1)

In this section, we provide a comprehensive comparison of DMI with various baseline methods, as shown in Table 2.

Firstly, it can be observed that deep learning-based methods, e.g., GRU4Rec, consistently outperform the traditional popularity-based approach by a significant margin. This observation is not surprising, since deep learning models can capture complex sequential patterns and non-linear relationships in users' historical interactions that simple popularity-based methods cannot model.

Secondly, the earlier multi-interest models such as MIND and ComiRec do not consistently outperform the single-vector models such as GRU4Rec and Y-DNN. This is likely due to their suboptimal training schemas, absence of temporal information, and inadequate training among multi-interest vectors. Subsequently, PIMIRec and RE4 enhanced the initial multi-interest learning models. RE4 focuses on developing distinct multi-interest vectors, while PIMIRec incorporates time series data. However, these approaches rely on the established multi-interest extraction framework from ComiRec or MIND, without advancing the framework itself. In contrast, REMI addresses the inherent problems of multi-benefit extraction frameworks, such as insufficient training efficiency and route collapse, achieving peak performance.

Lastly, our proposed DMI outperforms all baselines across all metrics and datasets, confirming its effectiveness in enhancing interest representations in multi-interest recommendation. Besides, DMI reveals the prevalent issue of interest noise in interest modeling, suggesting that the application of Diffusion in this context holds significant promise.

Table 3: Results of ablation study. Here, we report metric@50 on the Book and Gowolla datasets.

| Versions | Book | | | Gowolla | | |
|---|---|---|---|---|---|---|
| | Recall | NDCG | Improv. | Recall | NDCG | Improv. |
| DMI-diff | 0.1189 | 0.0657 | — | 0.2487 | 0.2116 | — |
| DMI-GD | 0.02891 | 0.0181 | -76.69% | 0.0516 | 0.0651 | -79.26% |
| DMI-T | 0.1235 | 0.0671 | 3.81% | 0.2670 | 0.2249 | 7.35% |
| DMI-IP | 0.1352 | 0.0733 | 13.70% | 26.93 | 0.2251 | 8.83% |
| DMI | **0.1402** | **0.0751** | **17.91%** | **0.2735** | **0.2263** | **9.97%** |

## 4.3 Ablation Study (RQ2)

In this section, we conduct comprehensive ablation studies to demonstrate the effectiveness of our key designs, including the denoising module, the item pruning strategy, the attention-based refinement, and the gradient detachment operation between the denosing module and the multi-interest extraction module. The ablation models include:

- **DMI-diff:** In DMI-diff, we remove the diffusion module and directly use the coarse-grained interest vectors from the multi-interest recommender.
- **DMI-T:** In DMI-T, we replace the Transformer layer with an MLP in the denosing module.
- **DMI-IP:** In DMI-IP, we do not select TopK relevant historical item embeddings through item pruning strategy. Thus, all historical item embeddings are attended in the reverse process, which could introduce interest-irrelevant information.
- **DMI-GD:** In DMI-GD, we do not stop the gradient from the diffusion module to the multi-interest recommender. The denoised vectors will be directly used as the final interest vectors without fusing with the initial interest vectors from the multi-interest recommender as in Eq. 13.

The results are shown in Table 3. From this table, we can observe that: (1) DMI outperforms all the variants without certain components, e.g., DMI-IP, DMI-T, and DMI-GD, confirming the effectiveness and the complementary nature of our key modules. (2) DMI-T performs worse than DMI, which implies the significance of using cross-attention mechanism to incorporate collaborative information. While MLP is widely used in diffusion recommenders for its simplicity and efficiency, it lacks effective guidance in the denoising process. (3) DMI significantly outperforms DMI-IP. This result is not surprising, since indiscriminately considers all historical behaviors as collaborative information will introduce potential noise. (4) In DMI-GD, we observe that the model struggles to converge normally. This highlights the difference between recommendation tasks (which align user and positive sample representations) and generative tasks (which focus solely on user modeling). In this context, a well-designed setup is crucial for balancing convergence and generative benefits.

Table 4: Category statistics analysis on the Book dataset.

| Method | Evaluate Metric@50 | | |
|---|---|---|---|
| | Conc ↓ | Div(all) ↑ | Div(hit) ↑ |
| REMI | 1.67 | 4.63 | 279 |
| DMI | 1.62 | 4.57 | 312 |
| Δ | -3.0% | -1.3% | +11.8% |

## 4.4 Impact of DMI on Matching Results (RQ3)

To better understand the effectiveness of DMI in generating fine-grained and focused interest vectors, we conducted several statistical experiments on the categories of the retrieved items. Specifically, we utilized the Amazon Book dataset and obtained the categories of each book ID. Then, we analyze the impact of DMI on the distribution of categories in the following experiments:

- **Category concentration** reflects the degree of concentration in the retrieved set. During experiments, we first retrieve corresponding items from the item corpus using both the refined vectors and the original vectors. Then, we calculate the Hamming distance between the categories of two items **x** and **y** in the

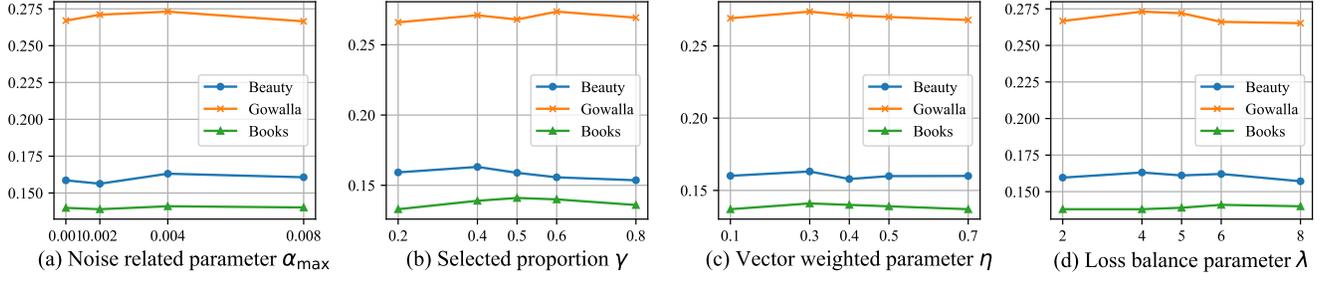

Figure 3: Analysis on different parameters. We show Recall@50 on three datasets.

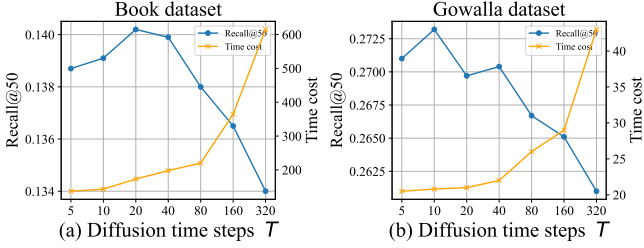

Figure 4: Performance and inference time cost comparison of using different diffusion steps in two datasets.

retrieved set to evaluate the degree of concentration:

$$d_H(\mathbf{x}, \mathbf{y}) = \sum_{i=1}^{n} \delta(x_i \neq y_i) \quad (21)$$

where $x_i$ and $y_i$ are the $i$-th elements of $\mathbf{x}$ and $\mathbf{y}$, respectively. The indicator function $\delta(x_i \neq y_i)$ equals 1 if $x_i \neq y_i$, and 0 otherwise. The final concentration score can be calculated as:

$$\text{Conc @ N} = \frac{\sum_{j=1}^{N} \sum_{k=j+1}^{N} d_H\left(\text{CATE}\left(\hat{i}_{u,j}\right), \text{CATE}\left(\hat{i}_{u,k}\right)\right)}{N \times (N-1)/2} \quad (22)$$

where CATE($i$) denotes the categories set of item $i$ and $\hat{i}_{u,j}$ denotes the $j$-th retrieved item for the user $u$.

- **Category diversity of items retrieved** In the matching stage, improving the diversity of recommendations is important. We follow the definition proposed by ComiRec [1] to calculate the diversity of the items retrieved using multiple interest vectors.

$$\text{Div(all) @ N} = \frac{\sum_{j=1}^{N} \sum_{k=j+1}^{N} \delta\left(\text{CATE}\left(\hat{i}_{u,j}\right) \neq \text{CATE}\left(\hat{i}_{u,k}\right)\right)}{N \times (N-1)/2} \quad (23)$$

- **Category diversity of items hit** We further analyze the categories of items that have been successfully interacted with by users in the model recommendation results. Specifically, the categories of items are calculated by summing the total number of distinct categories for each user's successfully interacted items.

As shown in Table 4, we can draw three key conclusions: (1) items retrieved by DMI exhibit higher category concentration, confirming the effectiveness of DMI in generating more fine-grained and focused interest representations. (2) The Div(all) metric shows a slight reduction in overall matching diversity. This result is expected, as some diversity may arise from interest-irrelevant components of the interest vectors. (3) A deeper analysis of Div(hit) metric reveals that DMI significantly enhances the diversity among successfully interacted items. This demonstrates that DMI learns more accurate interest vectors, which highly aligns with users' true preferences.

## 4.5 Hyper-parameter Study (RQ4)

*4.5.1 Inference Step Analysis.* The maximum diffusion step $T$ is a critical parameter controlling the noise level at the final stage of the diffusion process. It determines the retention of interest information before denoising and the number of inference steps. As shown in Figure 4, excessive steps are unnecessary, as optimal performance is often achieved with fewer steps. This is because recommendation tasks require suppressing excessive noise by tuning $\alpha_{\max}$ and $T$ to retain sufficient personalized information. When $\alpha_{\max}$ is appropriately set, smaller $T$ suffices, while larger $T$ disrupts personalization and causes reconstruction failure. Thus, DMI can achieve significant improvements with considerably less time and resource expenditure compared to traditional diffusion models.

Table 5: Performance of different interest heads number $k$ across three datasets. We show the metric@50 in the Table.

| $K$ | Book | | Gowolla | | Beauty | |
|---|---|---|---|---|---|---|
| | Recall | NDCG | Recall | NDCG | Recall | NDCG |
| 2 | 0.1317 | 0.0718 | 0.2595 | 0.2218 | 0.1551 | 0.0721 |
| 4 | 0.1402 | 0.0751 | **0.2731** | **0.2308** | 0.1631 | 0.0751 |
| 6 | 0.1413 | 0.0759 | 0.2667 | 0.2246 | 0.1576 | 0.0743 |
| 8 | **0.1420** | **0.0760** | 0.2685 | 0.2241 | **0.1665** | **0.0764** |

*4.5.2 Analysis on the Number of Interest Heads K.* We demonstrate the performance of DMI under different interest numbers $K$ in Table 5. We can observe that DMI generally performs better initially as the number of interest heads increases, suggesting that it captures more diverse interests. This is linked to DMI's ability to generate fine-grained and concentrated interest distributions. However, the relationship is not strictly linear; the optimal number of interest heads may depend on the characteristics of the dataset. For example, in the Gowalla dataset, the optimal number of interests is 4, likely due to the social-based characteristic of items. Following the mainstream setting [25], we fix $K$ to be 4 for other studies.

*4.5.3 Analysis on other Key Hyper-parameters.* To provide a more comprehensive analysis, we experimentally tune several key hyper-parameters to explore their impact on model performance. The results of these experiments are illustrated in Figure 3. Based on our findings, we draw the following conclusions: 1) $\alpha_{\max}$ controls the maximum noise level at each step of the forward process. In the context of recommender systems, this value is typically set to a small value to avoid excessive disruption of the user-specific information embedded in the vectors. Compared to other hyperparameters, $\alpha_{\max}$ has a relatively minor impact on model performance, particularly in the book dataset. In practice, it is often tuned jointly with the diffusion step $T$ to achieve optimal results. 2) As the selection ratio $\gamma$ decreases, DDRM's performance initially declines, indicating that the primary noise is not caused by items with the lowest attention scores. However, as $\gamma$ continues to decrease, performance improves significantly. This suggests the presence of a substantial amount of unexplored interest-related attributes, which were previously overlooked by the existing multi-interest extraction network due to its item-level aggregation approach. 3) The vector weighted parameter $\alpha$ and loss balance parameter $\lambda$ are all used to balance generative tasks and discriminative tasks. In practice, it is crucial to ensure that neither parameter is set too large, as this could hinder the convergence of the core recommendation task. Subsequently, gradually increasing their values always yields additional benefits from the generative task.

**Table 6: Performance improvements on online A/B test. The results are percentage numbers with "%" omitted.**

| Method | #Engagement | #T-Clicks | #T-Exposed | #Favorite | #Comment |
|---|---|---|---|---|---|
| DMI | +0.59 | +0.31 | +0.22 | +1.09 | +0.64 |

## 4.6 Online Result (RQ5)

In this section, we further test DMI to justify its effectiveness in real-world recommender systems. Specifically, we conduct an online A/B test over one week by deploying DMI in the mainfeed recommender system of [PLACEHOLDER] [3], which serves the major traffic of hundreds of millions of daily active users. To reduce resource consumption during deployment, we employed DDIM [23] to significantly reduce the steps required during inference sampling. It is also worth noting that the baseline system comprises multiple retrieval models. To ensure a fair comparison, we solely replace the multi-interest model (e.g., ComiRec) solely with DMI, while maintaining all other retrieval models unchanged.

The results are shown in Table 6. The main metric for this test is user engagement, measured by interaction metrics. It can be seen that DMI significantly outperforms the baseline system, achieving 0.59% improvement in user engagement, a typical metric reflecting user satisfaction. In the context of our mainfeed traffic, even a 0.2% improvement in user engagement is considered statistically significant and represents a meaningful enhancement in performance. Additionally, DMI enhances recommendation diversity, resulting in a 0.22% increase in the number of categories exposed and a 0.31% rise in categories clicked. These results verify DMI's superiority in capturing user preferences and enhancing overall system performance, demonstrating its potential as a general approach to improve user experience in the mainfeed.

---
[3]To comply with anonymity requirements, the specific name will be disclosed upon paper acceptance.

## 5 RALATED WORKS

### 5.1 Multi-Interest Learning

Multi-interest recommendation aims to capture the diverse interests of users by modeling complex interaction patterns, which overcomes the limitations of representing user preferences with a single vector. By enhancing diversity, a critical factor in the matching stage, multi-interest recommendation has attracted significant attention. MIND[11] first adopts a dynamic routing mechanism to aggregate users' historical behaviors into multiple interest capsules. Afterward, ComiRec [1] further investigates multi-head attention-based multi-interest routing for capturing the user's diverse interests and introduces diversity controllable methods. PIMIRec [3] and UMI [2] extend this framework by incorporating time-based features and profiles. Re4 [27] enhances the model by considering the backward flow to refine the interest vectors. RimiRec [19] explicitly models the hierarchical structure of user interests. Recently, some works [6, 25] have been proposed to solve the problems of routing collapse and interest vector entangled that come with the existing multi-interest framework. In this paper, we pointed out another limitation of the multi-interest framework and decided to move the item-level interest capture to the dimensions level.

### 5.2 Diffusion Models in Recommendation

Inspired by the success of diffusion models in image synthesis [8, 20] and text generation [12, 13, 15], various studies have explored the potential of DMs in recommender systems. For instance, DiffRec incrementally[29] generates interaction predictions from partially corrupted historical interaction data. In [17], DM is employed to model complex distributions in multimodal recommendation systems, effectively smoothing deviations between modal features and collaborative signals. DreamRec [26] leverages DM to explore the underlying distribution of item space and generate the oracle items. In contrast to the above studies leveraging DM's generative capabilities, several DM-based recommender systems concentrate on improving denoising for recommendation systems (SR). DiffKG [9] introduces a KG filter that eliminates irrelevant and erroneous data by integrating collaborative signals and extracts high-quality signals from the aggregated representation of noisy knowledge graphs. [28] utilize collaborative information to guide the denoising model against noisy feedback. Overall, these diffusion models guide the process using general collaborative information, typically extracted by an encoder from the entire historical sequence. In comparison, our proposed DMI leverages interest-specific collaborative signals to focus on the reconstruction of personalized attributes.

## 6 CONCLUSION

In this paper, we reveal the problem that existing multi-interest models only capture overall item-level relevance, leading to coarse-grained interest representations that include irrelevant information. To address this issue, we present DMI, a Diffusion Multi-Interest model for interest refinement at the dimension level. Specifically, we first implement a base multi-interest recommender to capture diverse user interests from historical behaviors. Second, we develop a denoising module to generate more fine-grained and focused interest representations. To guide the denoising process, we introduce an item pruning strategy together with a cross-attention mechanism that incorporates the interest-relevant collaborative information. Extensive offline and online experiments demonstrate the effectiveness of DMI, particularly in generating personalized while diverse interest representations.

## Acknowledgments

To Robert, for the bagels and explaining CMYK and color spaces.